\documentclass[sigconf]{acmart}
\renewcommand\footnotetextcopyrightpermission[1]{} % removes footnote with conference information in first column
\usepackage{amsmath,amsfonts}
\usepackage{algorithm}
\usepackage{graphicx}
\usepackage{textcomp}
\usepackage{xcolor}
\usepackage[utf8]{inputenc}
\usepackage{listings}
\usepackage{float}
\usepackage{subfigure}
\usepackage{url}
\usepackage{multirow}
\usepackage{commath}
\usepackage{algpseudocode}
\usepackage{etoolbox}
\usepackage{varwidth}
\usepackage{tikz}
\usetikzlibrary{shapes,arrows}
\usepackage{colortbl}

\AtBeginDocument{%
  \providecommand\BibTeX{{%
    \normalfont B\kern-0.5em{\scshape i\kern-0.25em b}\kern-0.8em\TeX}}}

\copyrightyear{2021}
\acmYear{2021}
\setcopyright{none}

\acmConference[ArXiv Preprint]{ArXiv Preprint}{2021}{USA}

\acmBooktitle{ArXiv Preprint, 2021, Las Cruces, NM, USA}

\acmPrice{15.00}
\acmDOI{10.xxx/xx.xx}
\acmISBN{xx-x-xxx-xx-3/20/09}

\begin{document}

%%
%% The "title" command has an optional parameter,
%% allowing the author to define a "short title" to be used in page headers.
\title{PPT-SASMM: Scalable Analytical Shared Memory Model}
\subtitle{Predicting the Performance of Multicore Caches from a Single-Threaded Execution Trace}

%%
%% The "author" command and its associated commands are used to define
%% the authors and their affiliations.
%% Of note is the shared affiliation of the first two authors, and the
%% "authornote" and "authornotemark" commands
%% used to denote shared contribution to the research.
% \author{Ben Trovato}
% \authornote{Both authors contributed equally to this research.}
% \email{trovato@corporation.com}
% \orcid{1234-5678-9012}
% \author{G.K.M. Tobin}
% \authornotemark[1]
% \email{webmaster@marysville-ohio.com}
% \affiliation{%
%   \institution{Institute for Clarity in Documentation}
%   \streetaddress{P.O. Box 1212}
%   \city{Dublin}
%   \state{Ohio}
%   \postcode{43017-6221}
% }

\author{Atanu Barai}
\affiliation{Klipsch School of ECE\\
New Mexico State University\\
Las Cruces, NM 88003, USA}
\email{atanu@nmsu.com}

\author{Gopinath Chennupati}
\affiliation{Los Alamos National Laboratory\\
Los Alamos, NM 87545, USA}
\email{gchennupati@lanl.gov}

\author{Nandakishore Santhi}
\affiliation{Los Alamos National Laboratory\\
Los Alamos, NM 87545, USA}
\email{nsanthi@lanl.gov}

\author{Abdel-Hameed Badawy}
\affiliation{Klipsch School of ECE\\
New Mexico State University\\
Las Cruces, NM 88003, USA}
\authornote{Also affiliated with Los Alamos National Laboratory, Los Alamos, NM, USA.}
\email{badawy@nmsu.com}

\author{Yehia Arafa}
\affiliation{Klipsch School of ECE\\
New Mexico State University\\
Las Cruces, NM 88003, USA}
\email{yarafa@nmsu.com}

\author{Stephan Eidenbenz}
\affiliation{Los Alamos National Laboratory\\
Los Alamos, NM 87545, USA}
\email{eidenben@lanl.gov}

%%
%% By default, the full list of authors will be used on the page
%% headers. Often, this list is too long and will overlap
%% other information printed in the page headers. This command allows
%% the author to define a more concise list
%% of authors' names for this purpose.
\renewcommand{\shortauthors}{Atanu, et al.}

%%
%% The abstract is a short summary of the work to be presented in the
%% article.
\begin{abstract}
Performance modeling of parallel applications on multicore processors remains a challenge in computational co-design due to multicore processors' complex design. Multicores include complex private and shared memory hierarchies. We present a Scalable Analytical Shared Memory Model (SASMM). SASMM can predict the performance of parallel applications running on a multicore. SASMM uses a probabilistic and computationally-efficient method to predict the reuse distance profiles of caches in multicores. SASMM relies on a stochastic, static basic block-level analysis of reuse profiles. The profiles are calculated from the memory traces of applications that run sequentially rather than using multi-threaded traces. The experiments show that our model can predict private L1 cache hit rates with 2.12\% and shared L2 cache hit rates with about 1.50\% error rate.
\end{abstract}

%%
%% The code below is generated by the tool at http://dl.acm.org/ccs.cfm.
%% Please copy and paste the code instead of the example below.
%%
% \begin{CCSXML}
% <ccs2012>
%   <concept>
%       <concept_id>10010520.10010521.10010528.10010536</concept_id>
%       <concept_desc>Computer systems organization~Multicore architectures</concept_desc>
%       <concept_significance>500</concept_significance>
%       </concept>
%  </ccs2012>
% \end{CCSXML}

% \ccsdesc[500]{Computer systems organization~Multicore architectures}

%%
%% Keywords. The author(s) should pick words that accurately describe
%% of the work being presented. Separate the keywords with commas.

%%
%% This command processes the author and affiliation and title
%% information and builds the first part of the formatted document.
\maketitle

\section{Introduction}
With the emergence of Exascale computing and Moore's law coming to a halt, high core counts on multicore processors with complex and large cache hierarchies have become common. Such complicated designs come with several challenges~\cite{john:exascale}, such as the efficient use of available computing cycles, memory delays,  and modeling the performance of caches. Designers of parallel applications that run on multicores have to work hard to leverage this extensive computing power. One of the critical factors that determine a parallel application's performance on a multicore processor is the availability of data to the cores. One way to measure an application's data availability is through its cache utilization ability, which directly impacts runtime performance.

Modern processors have shared caches, which significantly impact the performance of an application in the form of data locality and inter-process communication. These factors are both complex to analyze and hardware dependent. Simulation as a modeling tool helps understand and predict applications' behavior and study the impact of the above factors on performance in a multicore configuration. Co-design, which we define as modeling both hardware and software, helps to tune an application's performance. Most of the efforts in co-design have focused on getting simulation data from cycle-accurate dynamic instrumentation tools~\cite{Moguls,Davis-max-cmp-thpt,Ekman-perf-pwr,Huh-des-sp}. However, these simulations require a large number of runs and experimentation with many hardware configurations. Such configurations include variations in cache hierarchies, core counts, and problem sizes, all of which contribute to increasing design space complexity. Using cycle-accurate dynamic simulators to evaluate and predict performance does not scale well. Our solution is to build a scalable simulation model that relies on a detailed cache hierarchy model and application.

In analyzing a cache's performance, \textit{Reuse Distance Analysis}~\cite{Mattson:RD:IBM} is one of the commonly used techniques. Reuse distance is defined as the number of unique memory references between two references to the same memory reference. For sequential programs, reuse analysis is architecture-independent, whereas for parallel programs that run on multicores, reuse distance dependents on how the memory references of threads interact. Therefore, on multicores, \textit{Concurrent Reuse Distance} (CRD) profiles~\cite{Multicore:ding2009a} use a global stack to quantify reuse across thread-interleaved memory references, and thus accounts for data sharing and interaction between threads accessing shared caches. However, CRD profiles are unscalable as the core count increases, and the thread interactions increase; thus, the memory traces get large, which significantly changes the CRD profiles. On the other hand, \textit{Private-stack Reuse Distance} (PRD) profiles depend on how the tasks are scheduled among multiple cores.

In this paper, we introduce the Scalable Analytical Shared Memory Model (SASMM). SASMM relies on the prediction capabilities of the recently open-sourced Performance Prediction Toolkit (PPT)~\cite{ppt}. SASMM is based on reuse distance estimation methods. Our crucial innovation is to include the realistic scenario of caches shared among multiple threads of an application, compared to existing cache models even in the PPT library. SASMM  estimates shared and private cache hit rates in a multi-thread and complex cache hierarchy architecture for different applications, which the user can specify. We use a translator based on the {\em Rose} compiler~\cite{Rose_Compiler:Liao} to get the threaded version of a parallel code written in OpenMP~\cite{openmp}. We develop a compiler-driven technique to identify the threaded programs' basic blocks in measuring the exact probabilities of executing a given basic block of a program. We collect LLVM basic block~\cite{Lattner:LLVM} labeled memory trace from a sequential execution of translated code only once. Using this memory trace, we explore through different scheduling and interleaving strategies of execution to mimic the behavior of multi-threaded programs on shared-memory multicores. These strategies are carried out at the basic block level. We collect a basic block labeled memory trace generated from the translated program's sequential run and apply a probabilistic analytical method to measure both the \textit{PRD} and the \textit{CRD} profiles. Using these profiles, we measure cache hit rates of the applications. We evaluate our approach with the hit rates collected using the Cachegrind tool from Valgrind~\cite{valgrind}. The results show that the model accurately predicts cache hit rates compared to hit rates collected using the Cachegrind tool.

\section{Background}
\subsection{Execution of Parallel Application: Fork Join Model}
OpenMP uses fork-join model for parallel execution of a program. The program begins as a sequential application with a master thread. When the first parallel region construct is encountered, the master thread forks a team of almost identical parallel threads. The forked threads have access to all the variables from the master thread, and those are shared variables. These threads may also have private variables of their own and can identify themselves with unique thread number. When the threads finish executing all the parallel region statements, they synchronize and terminate (join), leaving only the master thread. It is also possible to have nested parallelism where one in the team of threads can fork recursively until it reaches a certain task granularity.

\subsection{Reuse Distance Analysis}
Reuse distance (D) of a memory address, also known as LRU stack distance, is the number of unique memory references made by a program between two consecutive references to the same address. Note that, when a memory address is referenced for the first time, D's reuse distance is $\infty$. Reuse profile is the histogram of reuse distances for all memory references of a program. Reuse distance analysis measures the locality~\cite{locality:Ding:2003:PWL, locality:Zhong:2009:PLA} of an application, which can be used to predict the cache performance of that application~\cite{performance:Beyls:RD:2001, performance:Sen:2013:ROM, performance:CaBetacaval:2003:ECM} and make cache management policy decisions~\cite{C.Management:Duong:2012:ICM}. For a fully associative cache with capacity C, a memory reference's reuse distance will always trigger a cache miss, if D $\geq$ C. Table~\ref{fig:RD_theory} shows the reuse distance calculation for a sample trace. In the example, $50\%$ of memory references will cause a compulsory cache miss. If we consider that cache size is three, then 13\% of all memory references will cause a capacity cache miss. In our work, we calculate the reuse profile at cache line granularity. The addresses we consider to calculate \emph{D} are cache line addresses.

\begin{table}[t!]
    \centering
    \caption{Reuse Distance Example}
    \begin{tabular}{cc}
    % \toprule
    \multicolumn{2}{c}
    {
        \centering
        \begin{tabular}{|c|cccccccc|}
           % Time & 1 & 2 & 3 & 4 & 5 & 6 & 7 & 8 \\
            \hline
            Address & a & b & a & c & b & d & d & a\\
            \hline
            {\text { RD }} & {$\infty$} & {$\infty$} & {1} & {$\infty$} & {2} & {$\infty$} & {0} & {3}\\
            \hline
        \end{tabular}
    } \\
%   \if 
%     \\
    
%         {
%             \setlength{\tabcolsep}{2pt}
%             \begin{tabular}{c|cccc|c}
%                 Distance &  0 & 1 & 2 & 3 & -1\\
%                 \hline
%                 Frequency & 1 & 1 & 1 & 1 & 4 \\
%                 \hline
%                 P(RD) & {1\(/ 8\)} & {1\(/ 8\)} & {1\(/ 8\)} & {1\(/ 8\)} & {4\(/ 8\)}
%             \end{tabular}
%         }
%         & 
%         \includegraphics[width=30mm,scale=0.15]{Reuse_Profile.png}
%     \fi
    \end{tabular}{}
    \label{fig:RD_theory}
\end{table}

Reuse distance analysis is robust and architecture-independent for sequential applications. The same reuse profile can be used to determine the performance of different cache sizes. This saves a significant amount of time in cache hit rate analysis as we do not have to collect memory traces for different cache configurations. Many attempts~\cite{Ding:2001:RDA, Berg:SS, Steen:UGhent} demonstrated the use of memory traces for reuse profile calculations. These approaches use binary instrumentation tools to collect memory traces. The memory traces used in most of these attempts are significant in size and time-consuming to process, thereby unscalable. However, recent attempts from Chennupati et al.~\cite{ppt-amm,chennupati:pmbs,chennupati:pads} demonstrated analytical models that scale with a small input run of a program. These attempts help predict the performance of an application on single-threaded programs. In a similar spirit, we model the private and shared cache performance of multicore programs.

% \begin{figure}[htbp]
%     \begin{tabular}{cc}
%     \multicolumn{2}{c}
%     {
%         \centering
%         \begin{tabular}{c|cccccccc}
%             Time & 1 & 2 & 3 & 4 & 5 & 6 & 7 & 8 \\
%             \hline
%             Address & a & b & a & c & b & d & d & a\\
%             \hline
%             {\text { RD }} & {$-1$} & {$-1$} & {1} & {$-1$} & {2} & {$-1$} & {0} & {3}
%         \end{tabular}
%     } \\
%     \\
    
%         {
%             \setlength{\tabcolsep}{2pt}
%             \begin{tabular}{c|cccc|c}
%                 Distance &  0 & 1 & 2 & 3 & -1\\
%                 \hline
%                 Frequency & 1 & 1 & 1 & 1 & 4 \\
%                 \hline
%                 P(RD) & {1\(/ 8\)} & {1\(/ 8\)} & {1\(/ 8\)} & {1\(/ 8\)} & {4\(/ 8\)}
%             \end{tabular}
%         } 
%         & 
%         \includegraphics[width=30mm,scale=0.15]{Reuse_Profile.png}
%     \end{tabular}{}

%     % \qquad
    
%     % \begin{tabular}{c|cccc|c}
%     %     Distance &  0 & 1 & 2 & 3 & $\infty$\\
%     %     \hline
%     %     Frequency & 1 & 1 & 1 & 1 & 4 \\
%     %     \hline
%     %     P(RD) & {1\(/ 8\)} & {1\(/ 8\)} & {1\(/ 8\)} & {1\(/ 8\)} & {4\(/ 8\)}
%     % \end{tabular}
%     % \newline
%     % \textcolor{red}{compulsory miss = 50\%}
    
%     \caption{Example of RD Profile}
%     \label{fig:RD_theory}
% \end{figure}

\subsection{Multicore Reuse Distances}
%Previous multicore RD research has revolved around developing techniques for acquiring profiles dynamically and verifying the accuracy, which is not scalable. 
Most of the multicore processors contain both shared and private caches. Although the locality of references of a parallel program in a multicore processor is somewhat architecture-specific, it largely depends on the application's characteristics. The corresponding thread of a core accesses the private cache while the shared cache is accessed through all the cores. Two separate reuse profiles, \textit{Concurrent} and \textit{Private-stack} reuse profiles (CRD and PRD) are used to model shared and private caches~\cite{Jiang:RD-Applicable-on-chip} respectively. We can interleave memory references from all cores on a single LRU stack to measure concurrent reuse profiles. This interleaving causes different types of interaction: \textit{dilation, overlap, and interception}~\cite{Wu-multicore-journal}. Table~\ref{fig:CRD_theory} shows the memory references from two cores. For access of \textbf{a} at time 4, CRD is two where its PRD is 1. Here CRD is larger than PRD, which shows \textit{dilation}. On the other hand, data sharing reduces dilation. For the memory reference of \textbf{a} at time 9, CRD is three, although there are four memory references between two consecutive memory references at times 4 and 9. This shows \textit{overlapping} as \textbf{d} is accessed by both cores inside reuse interval of \textbf{a}. Again for the reference \textbf{b} at time 10, the reused data itself is shared. So its CRD is two, which is less than its PRD.

\newcolumntype{P}[1]{>{\centering\arraybackslash}m{#1}}
\begin{table}[t!]
    \centering
    \caption{Concurrent Reuse Distance Example}
    \begin{tabular}{|P{2.5cm}|cccccccccc|}
        \hline
        Time       & 1 & 2 & 3 & 4 &  5 &  6 & 7 & 8 & 9 & 10\\
        \hline
        Core $C_1$ & a &   & b & a &  e &  &   & d & a & b\\
        \hline
        Core $C_2$ &   & c &   &   &   & d & b &   &   &  \\
        \hline
        Shared Memory Access & a  & c & b  & a  & e  & d & b & d  & a  & b \\
        \hline
    \end{tabular}
    \label{fig:CRD_theory}
\end{table}

Several recent works have focused on CRD profile and performance prediction of the shared cache~\cite{Multicore:Performance_metrics:Ding2014, Multicore:Modeling_Superscalar_Memory-Level_Parallelism, Multicore:Modeling_CMP_Cache_Capacity:Shi, Multicore:Miss_Rate_Prediction:Zhong, Multicore:Formalizing_Data_Locality:Ceballos}. Recently researchers attempted to use an analytical model and sampling to speed up the performance prediction~\cite{Jiang:RD-Applicable-on-chip, Multicore_Reuse_Analytical:Jasmine, Schuff:2010:AMR:1854273.1854286, multicore:stat_multiprocessor_cache:Berg, Multicore-Aware-Derek}. All these models require trace collection from parallel execution of an application for different numbers of threads. On the other hand, our model collects trace once from the sequential run of the application. From that trace, we predict shared cache performance for a different number of threads. This makes our model highly scalable with core counts.

\section{PPT-SASMM: Scalable Analytical Shared Memory Model}
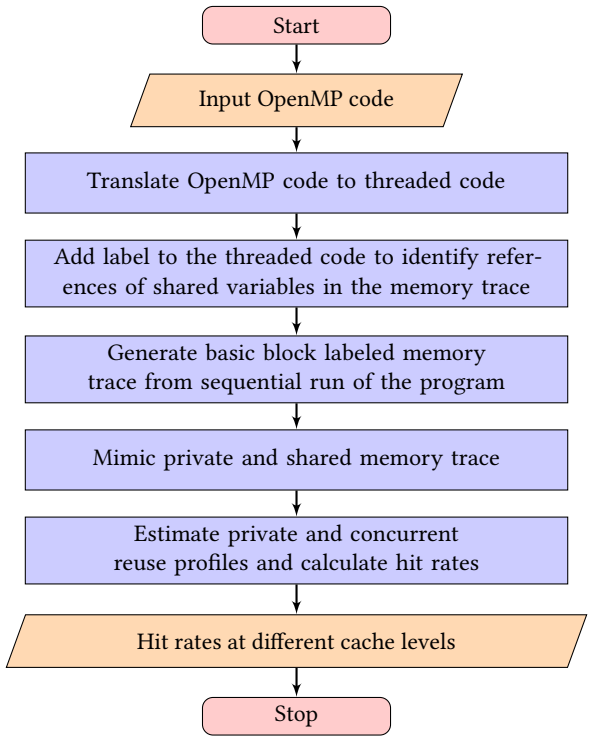
\begin{figure}[tb!]
\centering
\tikzstyle{startstop} = [rectangle, rounded corners, minimum width=2.5cm, minimum height=0.5cm,text centered, draw=black, fill=red!20]
\tikzstyle{io} = [trapezium, trapezium left angle=70, trapezium right angle=110, minimum width=2.3cm, minimum height=0.7cm, text centered, draw=black, fill=orange!30]
\tikzstyle{process} = [rectangle, minimum width=2.5cm, minimum height=0.8cm, text centered, text width=7cm, draw=black, fill=blue!20]
\tikzstyle{decision} = [diamond, draw, fill=, 
    text width=4.5em, text badly centered, node distance=3cm, inner sep=0pt]
\tikzstyle{block} = [rectangle, draw, fill=blue!20, 
    text width=5em, text centered, rounded corners, minimum height=4em]
\tikzstyle{line} = [draw, thick, -latex']
    
\begin{tikzpicture}[node distance = 1 cm]
    % Place nodes
    \node [startstop] (init) {Start};
    \node [io, below of=init] (input) {Input OpenMP code};
    \node [process, below of=input, yshift=-0.99mm] (translate) {Translate OpenMP code to threaded code};
    \node [process, below of=translate, yshift=-2.05mm] (labeladd) {Add label to the threaded code to identify references of shared variables in the memory trace};
    \node [process, below of=labeladd, yshift=-2.75mm] (tracegen) {Generate basic block labeled memory trace from sequential run of the program};
    \node [process, below of=tracegen, yshift=-2.05mm] (mimic) {Mimic private and shared memory trace};
    \node [process, below of=mimic, yshift=-2.05mm] (calulation) {Estimate private and concurrent reuse profiles and calculate hit rates};
    \node [io, below of=calulation, yshift=-2.05mm] (output) {Hit rates at different cache levels};
    \node [startstop, below of=output] (stop) {Stop};
    % Draw edges
    \path [line] (init) -- (input);
    \path [line] (input) -- (translate);
    \path [line] (translate) -- (labeladd);
    \path [line] (labeladd) -- (tracegen);
    \path [line] (tracegen) -- (mimic);
    \path [line] (mimic) -- (calulation);
    \path [line] (calulation) -- (output);
    \path [line] (output) -- (stop);
\end{tikzpicture}
    \caption{Flow Chart of the Scalable Analytical Shared Memory Model (SASMM)}\label{fig:flowchart}
\end{figure}

The scalable analytical shared memory model is a parameterized model for the performance prediction of parallel codes. We leverage reuse distance analysis to determine a parallel program's multicore reuse profile that runs on multiple cores. The reuse profiles are later used to determine the hit rates at different cache hierarchies. Figure~\ref{fig:flowchart} shows different steps of the analytical shared memory model. Various steps of our model include \emph{a)} translating the OpenMP program to a threaded program, \emph{b)} adding labels for shared variables in the threaded program, \emph{c)} generating a memory trace from basic block labels, \emph{d)} mimicking shared and private memory traces, \emph{e)} estimating private stack and concurrent reuse profiles and hit rates. We describe each of these steps in detail as follows.

\definecolor{codegreen}{rgb}{0,0.6,0}
\definecolor{codegray}{rgb}{0.5,0.5,0.5}
\definecolor{codepurple}{rgb}{0.58,0,0.82}
\definecolor{backcolour}{rgb}{0.95,0.95,0.92}
 
\lstdefinestyle{mystyle}{
    backgroundcolor=\color{backcolour},   
    commentstyle=\color{codegreen},
    keywordstyle=\color{magenta},
    numberstyle=\tiny\color{codegray},
    stringstyle=\color{codepurple},
    basicstyle=\ttfamily\footnotesize,
    breakatwhitespace=false,         
    breaklines=true,                 
    captionpos=b,                    
    keepspaces=true,                 
    numbers=left,                    
    numbersep=4pt,                  
    showspaces=false,                
    showstringspaces=false,
    showtabs=false, frame=single,                  
    tabsize=2
}
\begin{figure}[b!]
    \centering
    \lstset{style=mystyle}
\begin{lstlisting}[language=C,numbers=none]
int main()
{
    int i, n;
    int sum = 0;
    n = 500;
    #pragma omp parallel for reduction(+:sum)
    for (i = 0; i < n; i++)
        sum = sum + i;
}

\end{lstlisting}
    \caption{An Example OpenMP Program}
    \label{fig:simple_openmp}
\end{figure}

\begin{figure}[t!]
    \centering
    \lstset{style=mystyle}
\begin{lstlisting}[language=C,,numbers=none]
#include "libxomp.h" 

struct OUT__1__7285___data 
{
  void *n_p;
  void *sum_p;
}
;
static void OUT__1__7285__(void *__out_argv);

int main(argc,argv)
int argc;
char **argv;
{
  int status = 0;
  XOMP_init(argc,argv);
  int i;
  int n;
  int sum = 0;
  n = 500000;
  struct OUT__1__7285___data __out_argv1__7285__;
  __out_argv1__7285__ . sum_p = ((void *)(&sum));
  __out_argv1__7285__ . n_p = ((void *)(&n));
  XOMP_parallel_start(OUT__1__7285__,&__out_argv1__7285__);
  XOMP_parallel_end();
  XOMP_terminate(status);
}

static void OUT__1__7285__(void *__out_argv)
{
shared_var_trace0: {} 
  int *n = (int *)(((struct OUT__1__7285___data *)__out_argv) -> n_p);
  int *sum = (int *)(((struct OUT__1__7285___data *)__out_argv) -> sum_p);
other_trace1: {}
  int _p_i;
  int _p_sum;
  _p_sum = 0;
  long p_index_;
  long p_lower_;
  long p_upper_;
  XOMP_loop_default(0, *n - 1,1,&p_lower_,&p_upper_);
  for (p_index_ = p_lower_; p_index_ <= p_upper_; p_index_ += 1) {
    _p_sum = _p_sum + p_index_;
  }
  XOMP_atomic_start();
   *sum =  *sum + _p_sum;
  XOMP_atomic_end();
  XOMP_barrier();
}
\end{lstlisting}
    \caption{Transformed OpenMP code using Rose compiler}
    \label{fig:simple_rose_code}
\end{figure}

\subsection{Program Translation}
\label{sec:prog_translation}
In the first step, we convert the OpenMP application to an intermediate threaded code using OpenMP translator in ROSE~\cite{Rose_Compiler:Liao} compiler. In the translation process, the parallel sections of the original code are transformed into intermediate threaded code. The translation is important in order to track the reuse distances of shared variables. With the high-level OpenMP code, measuring the reuse distances of shared variables is difficult. Therefore, the translated code helps in efficient reuse analysis, thereby the shared cache performance.  The threaded version of the code contains XOMP wrapper functions (generated from the Rose compiler), that call GNU OpenMP (GOMP) (when compiled with GCC) library functions. The parallel sections' private variables are translated as local variables in the code's corresponding threaded version. Each thread under execution runs the XOMP wrapper functions, where each thread allocates memory for the local variables. The threaded version of the code's functions receives pointers' structure as a parameter for the shared variables. At the beginning of these functions, all the members of those structures are assigned to locally declared pointers. We create separate \emph{labels} for these shared parts of the code, which is where the assignments happen so that the memory trace of the shared variables of the code are grouped in the corresponding basic block labels (described in section~\ref{sec:mem_trace}). Figure~\ref{fig:simple_rose_code} shows the transformed  code of the simple OpenMP code in Figure~\ref{fig:simple_openmp}. In the translated code in Figure~\ref{fig:simple_rose_code} the static function named \emph{OUT\_\_1\_\_7285} corresponds to the parallel section of the simple OpenMP code in Figure~\ref{fig:simple_openmp}. The shared variables are passed to this function using a pointer to the structure named \emph{\_\_out\_argv}. We put the assignment statements of shared variables under \emph{shared\_var\_trace0} label. In the memory trace, all the references under the \emph{shared\_var\_trace0} label are grouped together.

\subsection{Memory Trace Generation for Different Cache Hierarchies}
\label{sec:mem_trace}
In the second step, we generate LLVM basic block labeled memory trace of the translated threaded program. The LLVM IR of the source code consists of basic blocks, consisting of a single entry and a single exit point. In producing the trace, we execute the translated code sequentially for the parametrized program. We use LLVM based instrumentation to generate the basic block labeled memory trace of the translated program through sequential execution. In this memory trace the \textit{i\textsuperscript{th}} basic block \textit{(BB\textsubscript{i})} of the labeled trace contains all the memory addresses that are accessed as a result of executing the corresponding straight-line code of \textit{(BB\textsubscript{i})}. For each shared section, marked with a {\em label}, we gather the corresponding memory references of those shared sections from the trace. Sequentially using this memory trace result, we mimic the memory access behavior of the parallel program and thus generate the private memory trace on each thread under execution.

As OpenMP works within a fork-join model, the parallel section of the OpenMP code is executed at the same time on different cores. Each core has its copy of the parallel section of the code. Note that only the master thread executes the code's sequential part and the corresponding parallel section of the code. We mimic this behavior by making copies of each basic block of the parallel sections' memory references. Our mimicking strategy tries to replicate the memory trace of an OpenMP program on multiple cores. For example, if the parallel program uses $4$ cores, we make four copies of a basic block. We then add an {\em offset} to the memory addresses for each of the cores under execution except the core executing master thread. The basic blocks selected belong to the parallel region of the code. The offset is carried out on all memory references of a parallel region's basic blocks except for the shared variables' memory references. Some basic blocks (\emph{loop iterations}) under the parallel region are executed multiple times. They appear multiple times in the labeled memory trace. After adding offsets in the same way, we distribute the memory references belonging to these basic blocks evenly among all the cores. We choose the offset in such a way that the mimicked memory references do not match with the original memory references produced in the sequential execution. This mimicking strategy helps to show that the memory references belong to different cores.

\begin{algorithm}[htp]
    \begin{algorithmic}[1]
    \Procedure{$gen\_prvt\_trc$}{$all\_bb$, $trace$, $shared\_var\_refs$}
        \State $each\_core\_trace \gets [[]*num\_cores]$
        \State $all\_bb\_wins \gets get\_all\_bb\_windows(trace)$
        \For{$bb_i$ $\textbf{in}$ $all\_bb$}
            \State $bb_i\_wins \gets all\_bb\_wins[bb_i]$
            \State $len\_bb_i\_wins \gets \textbf{len}(bb_i\_wins)$
            \If {$bb_i$ $\textbf{in}$ $parallel\_bbs$}
                \If{$len\_bb_i\_wins == 1$}
                    \For{$core\_id$ $\textbf{in}$ $\textbf{range}\left(num\_cores\right)$}
                        \State each\_core\_trace[core\_id] $\gets$ \State\hspace{\algorithmicindent}$trace\left[bb_i\_wins\right]$
                    \EndFor
                \Else
                    \State $split\_wins \gets \textbf{array\_divide}(bb_i\_wins,$
                    \State\hspace{\algorithmicindent} $num\_cores, chunk\_size)$
                    \For{$core\_id$ $\textbf{in}$ $\textbf{range}\left(num\_cores\right)$}
                        \State each\_core\_trace[core\_id] $\gets $
                        \State\hspace{\algorithmicindent} $trace[split\_wins[core\_id]]$
                    \EndFor
                \EndIf
            \Else
                \State {$each\_core\_trace[0] \gets trace[bb_i\_wins]$}
            \EndIf
       \EndFor
    \EndProcedure
    \end{algorithmic}
    \caption{Private Memory Trace Generation}
    \label{alg:private_tr}
\end{algorithm}

The private caches (such as $L_1$) contain thread-specific execution where each core will have thread-specific memory trace. Therefore, we employ the procedure described in Algorithm~\ref{alg:private_tr} to generate private traces for each core, thereby calculating the corresponding reuse profiles and hit-rates. It takes a list of all the basic blocks, the sequential memory trace, and the references belonging to shared variables as input. It finds all instances of each basic block (\textit{BB$_i$}) in the memory trace and counts the number of instances. We refer to these instances as windows. If the basic block is in the parallel section with only one instance in the memory trace, we make a copy of that for each core, add offset to the memory references and assign it to each core. If that basic block has multiple instances in the memory trace, then we evenly distribute them to each core. We can perform distribution with chunk size, which is similar to OpenMP static scheduling chunk size. When the basic block is part of the code's sequential region, then we assign all the memory references of that basic block to the core executing the main thread. We find the list of basic blocks using our LLVM based offline code analysis tool.

The original OpenMP execution contains different scheduling strategies (static, dynamic, and guided) to execute the parallel sections. Recording memory traces for such scheduling strategies is cumbersome and inefficient in terms of both time and memory. Therefore, our model in this paper tries to generate a trace similar to the OpenMP scheduled traces. Here, we use the above recorded sequential trace to mimic the interleaving of threads. Our mimicking strategy distributes the corresponding memory threads equally among multiple threads under execution, similar to following static scheduling in OpenMP. We distribute the iterations to the cores according to an adaptive chunk size. In order to further study the effect of scheduling strategies on memory reuse, we propose various interleaving and scheduling strategies, described in section~\ref{sec:interleave}. %which is beyond the scope of this paper, and we reserve it for future work.

\begin{algorithm}[t!]
    \begin{algorithmic}[1]
    \Procedure{$interleave\_traces$}{$all\_bb$, $prvt\_mem\_traces$}
        \State $num\_of\_traces \gets \textbf{len}\left(prvt\_mem\_traces\right)$
        \State $shared\_mem\_trace \gets prvt\_mem\_traces[0]$
        \For{$bb_i$ $\textbf{in}$ $all\_bb$}
            \If {$bb_i$ $\textbf{in}$ $par\_bbs$}
                \For{$trace\_id$ $\textbf{in}$ $\textbf{range}(num\_trces)$}
                    \State $all\_trace\_bb_i\_wins[trace\_id] \gets$
                    \State\hspace{\algorithmicindent}$get\_bb_i\_windows(bb\_i, prvt\_mem\_$
                    \State\hspace{\algorithmicindent}$traces[trace\_id])$
                    \State $num\_bb_i\_instnc[trace\_id] \gets $ \State\hspace{\algorithmicindent}$\textbf{len}(all\_trace\_bb_i\_wins[trace\_id])$
                \EndFor
                \For{$instance$ $\textbf{in}$ $\textbf{range}(num\_bb_i\_instnc[0])$}
                    \State $ref\_instnc\_i\_all\_traces \gets [[]*num\_trces]$
                    \For{$tr\_id$ $\textbf{in}$ $\textbf{range}(num\_trces)$}
                        \State $ref\_instnc\_i\_all\_traces[tr\_id] \gets$ 
                        \State\hspace{\algorithmicindent}$prvt\_mem\_traces[tr\_id][all\_trace\_$
                        \State\hspace{\algorithmicindent}$bb_i\_wins[tr\_id][instance]]$
                    \EndFor
                    \State $tr\_id \gets 0$
                    \While {$ref\_instnc\_i\_all\_traces\; \neq []$}
                        \If{$strategy == uniform$} \label{line:startstrategy}
                            \State $tr\_id \gets randint(0, num\_trces - 1)$
                        \ElsIf{$strategy == round\_robin$}
                            \If{$tr\_id == num\_trces$}
                                \State $tr\_id \gets 0$
                            \Else
                                \State $tr\_id += 1$
                            \EndIf
                        \EndIf \label{line:endstrategy}
                        \State $interleaved\_bb_i\_trace \gets refs\_instnc\_$
                        \State\hspace{\algorithmicindent}$i\_all\_traces[tr\_id].pop(0)$
                    \EndWhile
                \EndFor
                \State $shared\_mem\_trace.replace\_bb_i\_refs(interlea$
                \State\hspace{\algorithmicindent} $ved\_bb_i\_trace)$
                
            \EndIf
       \EndFor
    \EndProcedure
    \end{algorithmic}
    \caption{Interleave memory traces}
    \label{alg:interleave_tr}
\end{algorithm}{}

For a shared memory trace, we take the {\em labeled} memory references from the basic block labeled private traces above. We interleave the memory references of the same basic block from all private traces of the cores sharing that particular memory. We try \emph{round-robin} and \emph{uniform random} scheduling to interleave the memory references. The resultant trace contains all basic blocks' memory trace under sequential execution and interleaved traces of all basic blocks under parallel execution. Thus, the sequence of basic blocks in the mimicked trace is retained from the sequential trace sequence. Similar traces can be generated with binary instrumentation tools such as \emph{Valgrind}~\cite{valgrind} and \emph{Pin}~\cite{pin-tool}. However, we use an LLVM based tool to leverage the conceptual advantage of dealing with simple straight line basic blocks within a program. Valgrind's \emph{Lackey} tool runs the multi-threaded program sequentially per thread, where the threads' interleaving is left to the operating system. Therefore the resultant memory trace happens to be multi-threaded.
On the other hand, with Pin, one has to produce a sequential trace and propose interleaving strategies. Nonetheless, we cannot derive a basic block labeled trace from Pin instead of our LLVM instrumentation. We estimate the reuse distances for each reference in the trace, once we have the memory trace that mimics the multicore execution.

\subsection{Interleaving Strategies}
\label{sec:interleave}
To analyze the performance on shared caches (such as $L_2$), we employ multiple interleaving strategies. This is to mimic the execution strategies of OpenMP constructs over the shared variables. Algorithm \ref{alg:interleave_tr} takes private traces as inputs and applies our interleaving strategies to generate shared traces for shared memory accesses. The algorithm inputs are a list of all the $BB_i$ and a two-dimensional list of private memory traces of all the various cores under consideration. Note that only the core executing the master thread has memory trace for the sequential section of the code and the trace for the parallel section. We assume that the master thread is being executed in \emph{core 0} without loss of generality. We initiate \emph{shared\_mem\_trace} with private memory trace of \textit{core 0}. In our next step, we find each basic block's \textit{$BB_i$} instances in all the private memory traces and count the number of instances for each trace. Then, we get the memory references and interleave the references for each instance of the private memory traces' basic block. We use either uniform-random or round-robin scheduling to interleave the traces. We replace the $n^{th}$ instance of \textit{$BB_i$} in \emph{shared\_mem\_trace} with corresponding interleaved \textit{$BB_i$} instance. The \textit{$BB_i$}s under sequential execution are not interleaved and remain unchanged. It is thus possible to mimic shared memory traces using this algorithm for any specific cache configuration.

We employ two interleaving strategies: {\em round-robin} and {\em uniform-random} (see lines~\ref{line:startstrategy}--\ref{line:endstrategy}). We employ these strategies on the sequential trace of a program, which in the end mimics the shared memory trace of a multicore program. For example, when we run a {\em for loop} for {\em 100} iterations, to mimic the trace of a $4$ core execution, we split the $100$ executions of each basic block of a {\em for loop} (note that a for loop, typically contains on the order of $5$ basic blocks) into $4$ parts, where each part belongs to a single core. We use the $4$ part trace to mimic the multicore trace, on which we employ the interleaving strategies. The two interleaving strategies is to experiment with different OpenMP scheduling strategies. In the {\em round-robin} strategy, for a given basic block, we take the memory reference from each of the four cores, that is core $0$, $1$, $2$, and $3$; then we repeat from core $0$ to $3$ for all the memory references of a basic block. In this way, the shared memory trace for $4$ cores is used to calculate the shared reuse profile across $4$ cores. In the {\em uniform-random} strategy, we select a number between $0$ and $3$ randomly using a uniform distribution. From that trace, we select a memory reference. We repeat this process until all the references from all $4$ parts of the trace are finished.
% We store memory trace of (\emph{core 0}) to another variable names \emph({shared_mem_trace}). Now, for each basic block \textit{BB$_i$} belonging to the parallel sections, we find their instances in each trace and store them as basic block windows.

\subsection{Calculating Basic Block Probabilities}
In the third step, we calculate the probability of executing each basic block from the basic block labeled sequential execution trace of the program. Let us assume that \textit{$BB_1$, $BB_2$, ..., $BB_j$, $BB_k$, ..., $BB_n$} are the basic blocks and any basic block can pass program execution flow to any other basic block. Let us also assume that these basic blocks are executed \textit{$N_1, N_2, ..., N_j, N_k, ..., N_n$} number of times respectively. Thus, the apriori probability of executing a basic block \textit{$P(BB_i)$} is:

\begin{equation}\label{eq:pbbi}
P(BB_i) = \dfrac{N_i}{\sum_{j=0}^{n} N_j}
\end{equation}

A particular basic block is executed depending on the number of occurrences of a basic block in the memory trace. We get the values of \textit{$N_i$} by counting the number of \textit{$BB_i$} instances in the memory trace. Note, \textit{$N_i$} changes with input size, as a result \textit{$P(BB_i)$} also changes. For sequential execution, these probabilities are valid for all levels of caches in the hierarchy. For parallel execution, each core uses a private cache along with shared caches to fetch data. Thus, for parallel execution, these probabilities are valid only for the last level cache. For private caches we calculate \textit{$BB_i$} probabilities using Eq.~\ref{eq:pri_pbbi} and \ref{eq:shar_pbbi}.

\begin{equation}\label{eq:pri_pbbi}
\operatorname{}\begin{array}{c}{P(BB_{ji}) }\\ { j \in serial}\end{array}=\frac{P(BB_{i})}{{\sum\limits_{k \in serial} P(BB_{k})} + {\sum\limits_{l \in parallel} P(BB_{l})}}
\end{equation}

\begin{equation}\label{eq:shar_pbbi}
\begin{array}{c}{P(BB_{ji}) }\\ {j \in parallel}\end{array}=\frac{\frac{P(BB_{i})}{m}}
{{\sum\limits_{k \in serial} P(BB_{k})} + {\sum\limits_{l \in parallel} \frac{P(BB_{l})}{m}}}
\end{equation}

\noindent where, \textit{$P(BB_i)$} denotes probability of the basic block under consideration, \textit{m} denotes the number of cores, \textit{$P(BB_k)$} denotes probability of a basic block of sequential region and \textit{$P(BB_l)$} denotes probability of a basic block of parallel region. We find the values of \textit{$P(BB_i)$, $P(BB_k)$ and $P(BB_l)$} from the memory trace of sequential execution. Eq~\ref{eq:pri_pbbi} is used for calculating probabilities of basic blocks of sequential region, while Eq~\ref{eq:shar_pbbi} is used for basic blocks of parallel region of the program.

\subsection{Probabilistic Reuse Profile Estimation}
In our next step, we analytically estimate the  \textit{Private-stack} and the \textit{Concurrent} reuse profiles of the program (P(D)) from our mimicked private and shared memory traces. The conventional methods of measuring the reuse profile are costly because of the enormous size of the memory traces. We use a technique described in~\cite{chennupati:pmbs}, which produces reuse distances at smaller input sizes of a program, and from those reuse distances, we estimate reuse profiles at more massive input sets. We estimate the reuse profile of a program using Eq.~\ref{eq:reuse_profile_program}.

\begin{equation}\label{eq:reuse_profile_program}
\operatorname{P}(D)=\sum_{i=0}^{n(B B)} P\left(B B_{i}\right) \times P\left(D | B B_{i}\right)
\end{equation}

\noindent where \textit{n(BB)} is the number of basic blocks, \textit{$P(BB_i)$} is the apriori probability of executing a basic block , \textit{D} is the reuse distance and \textit{$P(D|BB_i)$} is the conditional reuse profile of \textit{$i^{th}$} basic block.

\begin{algorithm}[t!]
    \begin{algorithmic}[1]
    \Procedure{$cond\_reuse\_prof\_BB_i$}{$bb$, $mem\_trace$}
        \State $reuse\_dists \gets [ ]$
        \State $sample\_size \gets x$
        \State $bb_i\_wins \gets get\_bb_i\_windows(bb\_i, trace)$
        \If{$len(bb_i\_wins) == 0$}
            \State $return$ $0$
        \EndIf
        \State $sampled\_windows \gets random(bb_i\_wins, sample\_size)$
        \For{$window$ $\textbf{in}$ $sampled\_windows$}
            \For{$idx, mem\_ref$ $\textbf{in enumerate}$ $window$}
                \State $rd\_val \gets get\_rd(idx, mem\_ref, mem\_trace)$ 
                \State $reuse\_dists.append(rd\_val)$
            \EndFor
        \EndFor
        \State $unique\_rds, counts \gets unique(reuse\_distances)$
        \State $p\_rd \gets {\bf map}({\bf lambda}$ $x$:$x/len(reuse\_dists)$, $counts)$ 
        \State $reuse\_prof \gets$ {\bf zip}$(uniq\_rds$, $p\_rd)$
    \EndProcedure
    \end{algorithmic}
    \caption{\textit{$P(D|BB_i)$} Calculation}
    \label{alg:cond_rp}
\end{algorithm}{}
\setlength{\textfloatsep}{0.1cm}
\setlength{\floatsep}{0.1cm}
% \vspace*{-.1cm}

\renewcommand{\arraystretch}{1.5}
\begin{table*}[htbp]
    \centering
    \caption{Applications used to verify our model. $\dagger$, $\ast$, and $\diamondsuit$ denote applications from PolyBench/OpenMP~\cite{polybench}, Rodinia/OpenMP~\cite{rodinia}, and PARSEC~\cite{parsec} benchmark suites respectively.}
    % \noindent\makebox[\textwidth]{
    \begin{tabular}{||c|P{3.5cm}|P{3.5cm}|c|c|P{1cm}||}
    \hline
    \textbf{Application} & \textbf{Description} & \textbf{Domain}  & \textbf{Input Size} & \textbf{Trace Size} & \textbf{Abbr.}\\
    \toprule
    \hline
    % \midrule
    $\dagger$ ADI & Alternating Direction Implicit method for 2D heat diffusion & Stencils & N=512, TSTEPS=2 & 2.1 GB & \textbf{adi}\\
    \hline
    $\ast$ BFS & Breadth-First Search & Graph Traversal & 64K Nodes & 1.1 GB & \textbf{bfs}\\
    \hline
    $\diamondsuit$ Blackscholes & Black-Scholes partial differential equation & Recognition, Mining and Synthesis & Options=4096, Runs=100 & 1.7 GB & \textbf{blk}\\
    \hline
    $\dagger$ Convolution-2D & 2D Convolution & Stencils & 1024 & 1.5 GB & \textbf{c2d}\\
    \hline
    $\dagger$ Durbin & Yule-Walker equations solver & Linear Algebra & 2048 & 4.2 GB & \textbf{dbn}\\
    \hline
    $\dagger$ Gramschmidt & QR decomposition with modified Gram Schmidt & Linear Algebra & 192 & 3.9 GB & \textbf{grm}\\
    \hline
    $\dagger$ Jacobi & Jacobi Iteration & Stencils & N=1024, Iterations=1024 & 3.0 GB & \textbf{jcb}\\
    \hline
    $\dagger$ LU & LU decomposition without pivoting & Linear Algebra & 256 & 3.2 GB & \textbf{lu}\\
    \hline
    $\dagger$ 2MM & Two Matrix Multiplication & Linear Algebra & 128 & 967 MB & \textbf{2mm} \\
    \hline
    \end{tabular}
    % }
    \label{table:benchmarks}
\end{table*}
% \clearpage

Algorithm~\ref{alg:cond_rp} calculates conditional reuse profile \textit{$P(D|BB_i)$} of a basic block. It takes the basic block and a mimicked memory trace as input, identifies the basic block's windows in memory trace, randomly selects \textit{sample\_size} windows. Typically we randomly select $1\%$ samples of basic block windows. For all memory addresses of each sampled window, we calculate its reuse distance, from which we calculate the corresponding probabilities. This sampling strategy saves significant time in the overall reuse distance calculation. Note that some basic blocks may not be executed at all in the program. In that case, there will be no window of that basic block in the memory trace.

\subsection{Hit Rate Estimation}
With the probabilistic \textit{Private-stack} and \textit{Concurrent} reuse profiles of each cache level, we measure private and shared cache hit rates using an analytical memory model, a stack distance based cache model (SDCM)~\cite{brehob:analytical}. Eq.~\ref{eq:phd} shows how to measure the hit rate at a given reuse distance ($P(h\mid D)$).
\begin{equation}\label{eq:phd}
P(h\mid D) =  \sum_{a=0}^{A-1}\binom{D}{a}\biggl(\dfrac{A}{B}\biggr)^a\biggl(\dfrac{B-A}{B}\biggr)^{(D-a)}
\end{equation}
\noindent where {\em D} is the reuse distance at cache line granularity, {\em A} is the associativity of the cache and {\em B} is cache size in terms of number of blocks (which is cache size over cache line size). Typically, Eq.~\ref{eq:phd} is used for an $n$-way associative cache. For a direct-mapped cache, probability of hit is defined as

\begin{equation}\label{eq:phd_direct}
P\left(h\mid D\right) = \textit{$\biggl(\frac{B-1}{B}\biggr)^{D}$}
\end{equation}

Finally, we calculate approximated unconditional the probability of a hit {\em P(h)} for the entire program as shown in Eq.~\ref{eq:phit}

\begin{equation}\label{eq:phit}
P(h) = \sum\limits_{i=0}^N P({D_i}) \times P({h\mid D_i})
\end{equation}

\noindent where, $P({D_i})$ is the probability of $i^{th}$ reuse distance ($D$) in a reuse distribution $Pr(D)$. These hit rates can be further used in runtime prediction of the applications, which is beyond this paper's scope.

%%%%%%%%%%%%%%%%%%%%%%%%%%%%%%%%%%%%%%%%%%%%%%%%%%%%%%%%%%%%%%%%%%%%%%%%%%%%%%%%

\begin{figure}[tbh!]
\centering
\subfigure[\textbf{L1 Hit Rates when applications run on 1-core CPU}]{
\label{fig:L1-1Core}%
\includegraphics[width=0.98\linewidth]{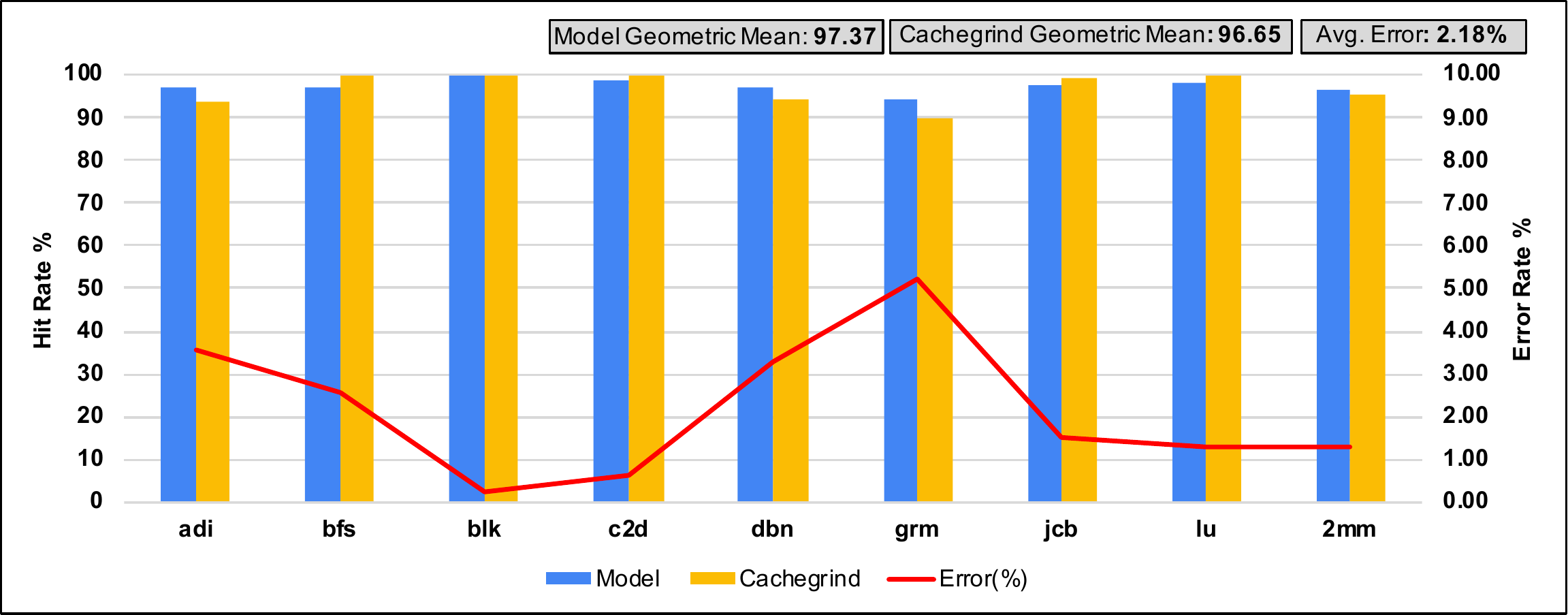}}%
\vspace{3pt}
\qquad
\subfigure[\textbf{L1 Hit Rates when applications run on 2-core CPU}]{
\label{fig:L1-2Core}%
\includegraphics[width=0.98\linewidth]{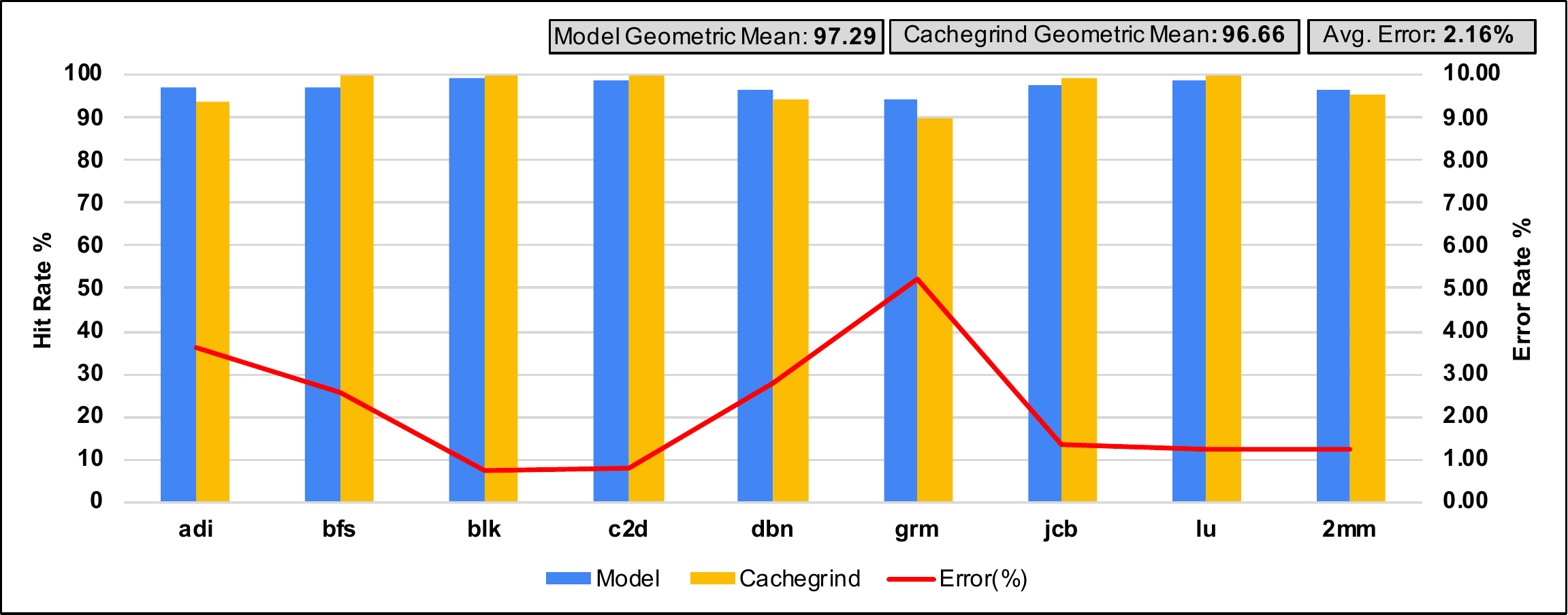}}%
%\vspace{3pt}
\qquad
\subfigure[\textbf{L1 Hit Rates when applications run on 4-core CPU}]{
\label{fig:L1-4Core}%
\includegraphics[width=0.98\linewidth]{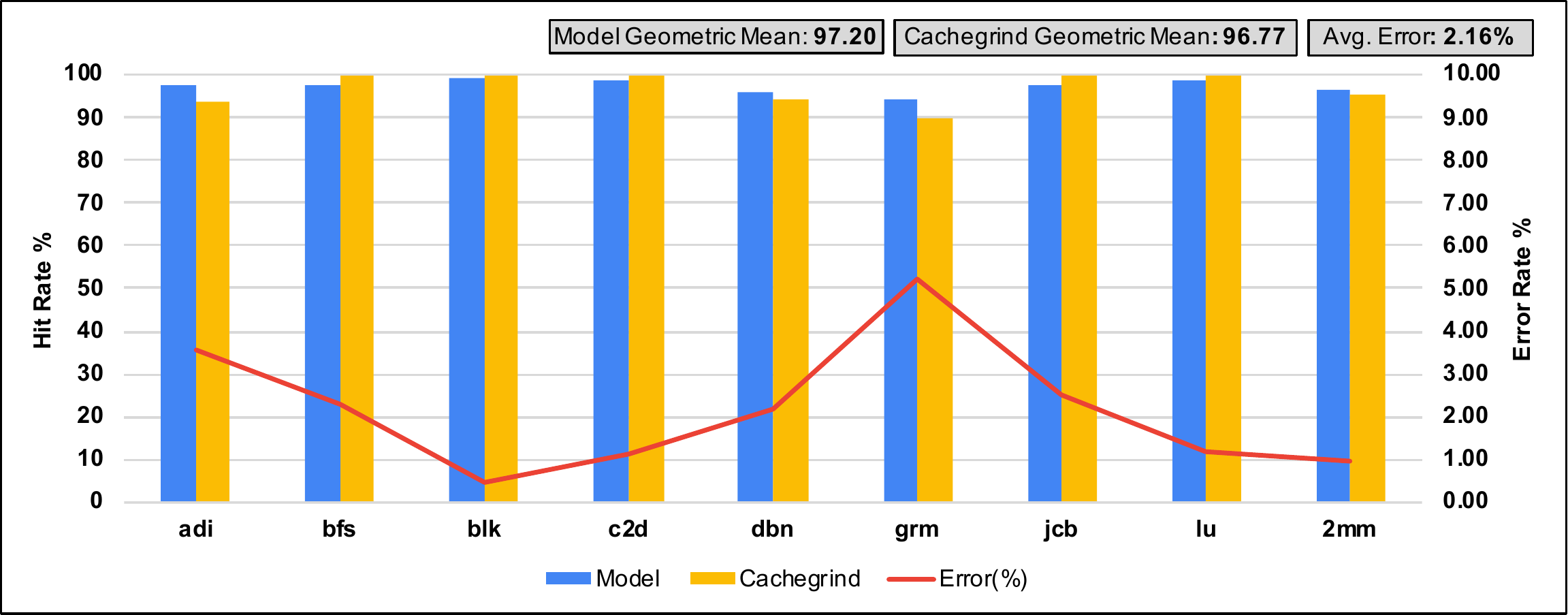}}%
%\vspace{3pt}
\qquad
\subfigure[\textbf{L1 Hit Rates when applications run on 8-core CPU}]{
\label{fig:L1-8Core}%
\includegraphics[width=0.98\linewidth]{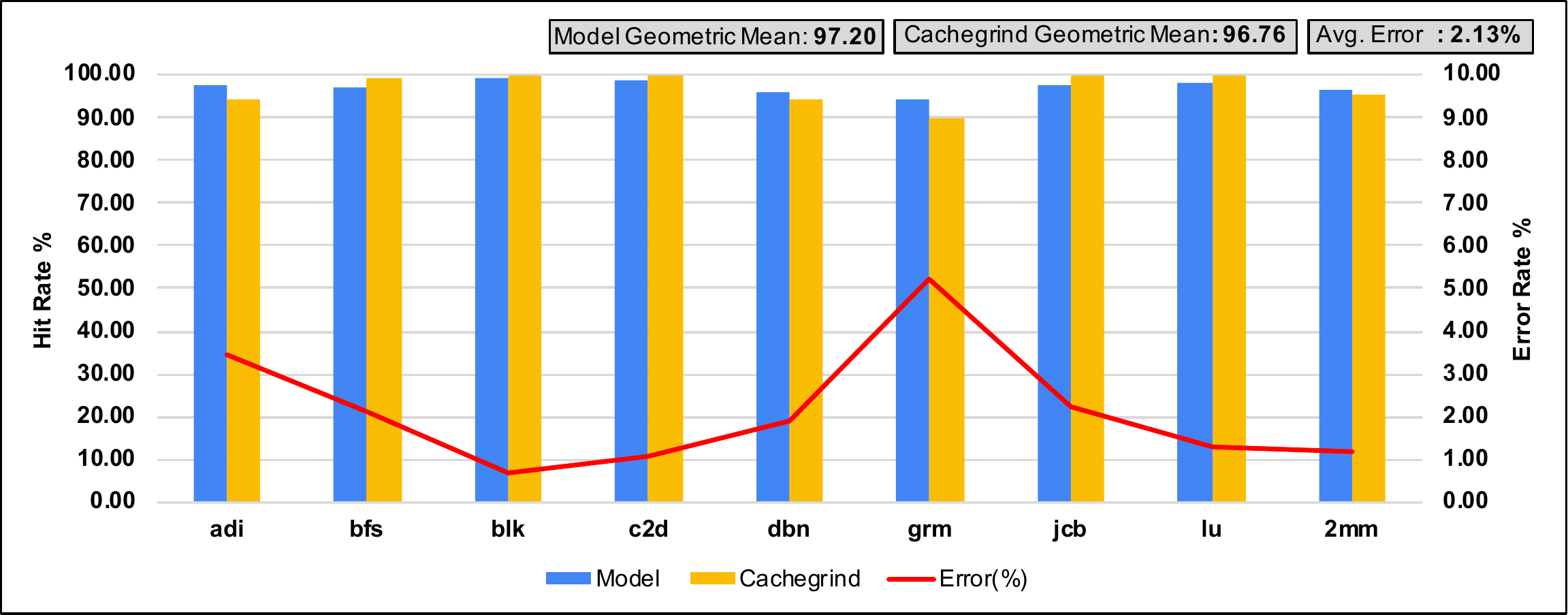}}%
\vspace{3pt}
\qquad
\subfigure[\textbf{L1 Hit Rates when applications run on 16-core CPU}]{
\label{fig:L1-16Core}%
\includegraphics[width=0.98\linewidth]{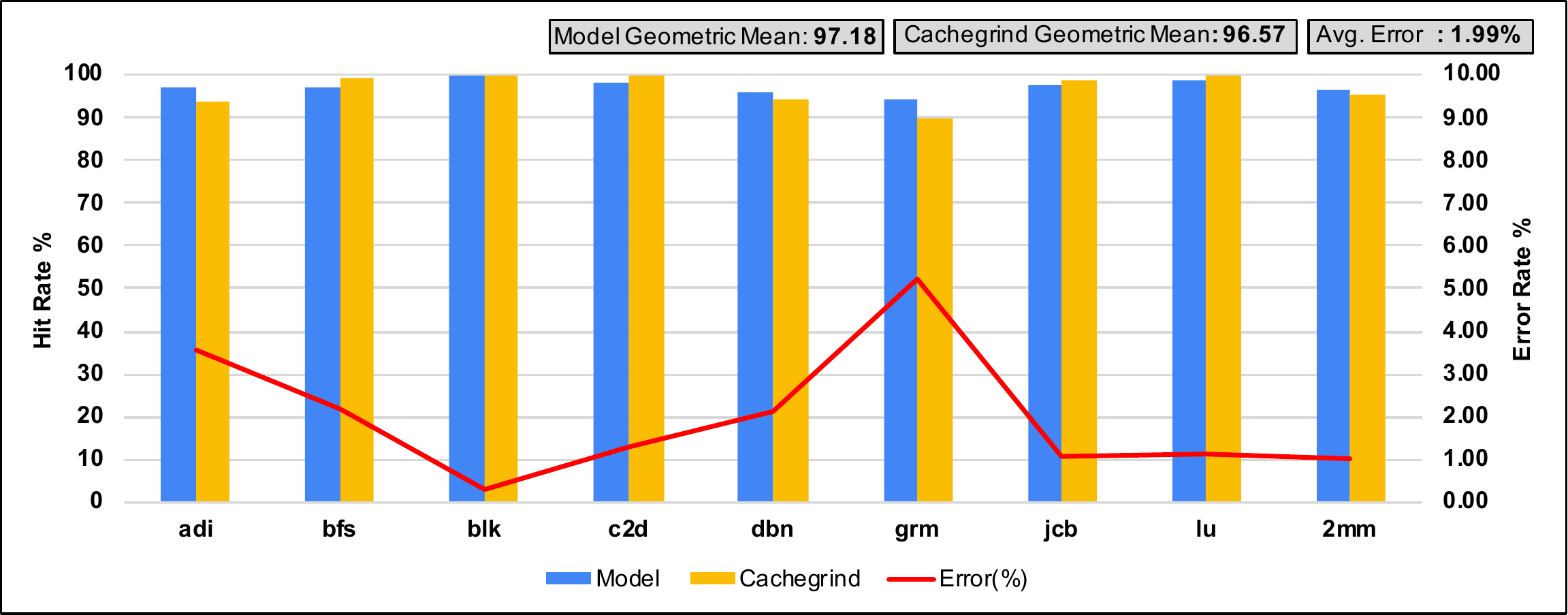}}%
\qquad
\caption{Hit Rate Comparison on 8KB Private L1-D Cache}
\label{fig:results-L1}%
\end{figure}

\begin{figure}[tbh!]
\centering
\subfigure[\textbf{L2 Hit Rates of applications running on single-core CPU (No Interleaving)}]{
\label{fig:L2-1Core}%
\includegraphics[width=0.98\linewidth]{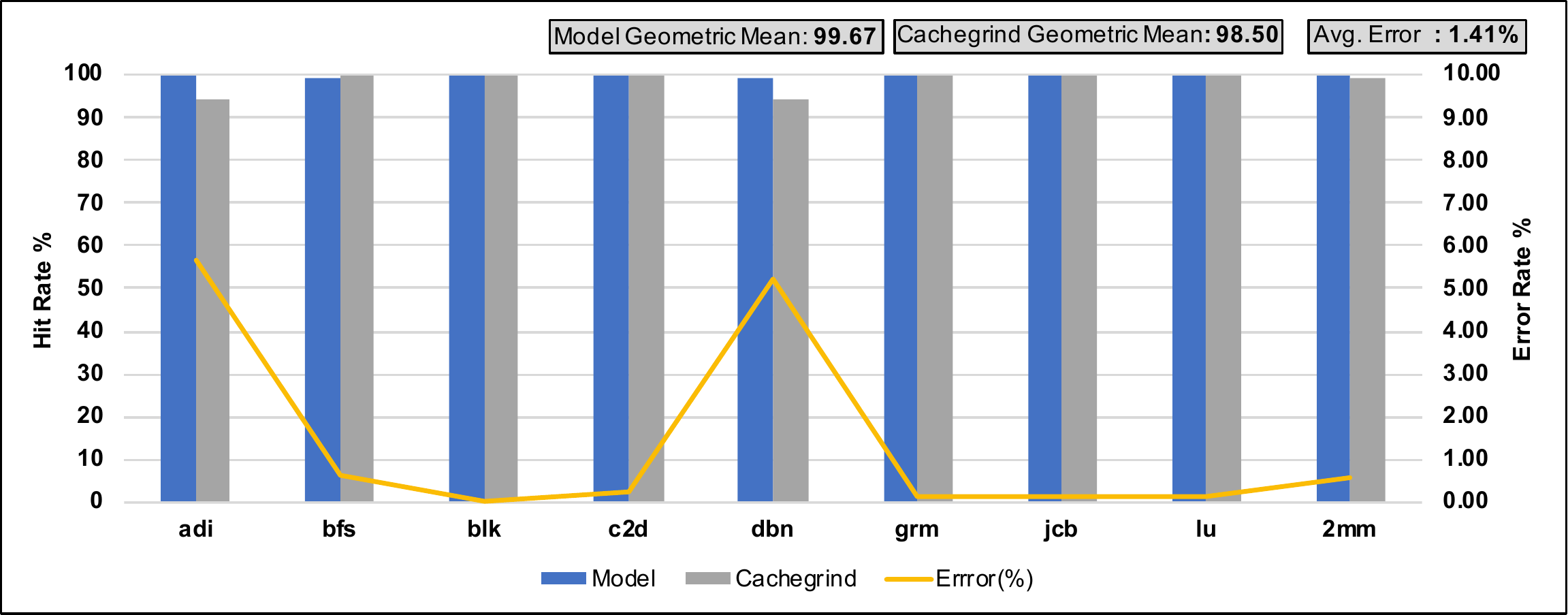}}%
\vspace{3pt}
\qquad
\subfigure[\textbf{L2 Hit Rates of applications running on 2 core CPU}]{
\label{fig:L2-2Core}%
\includegraphics[width=0.98\linewidth]{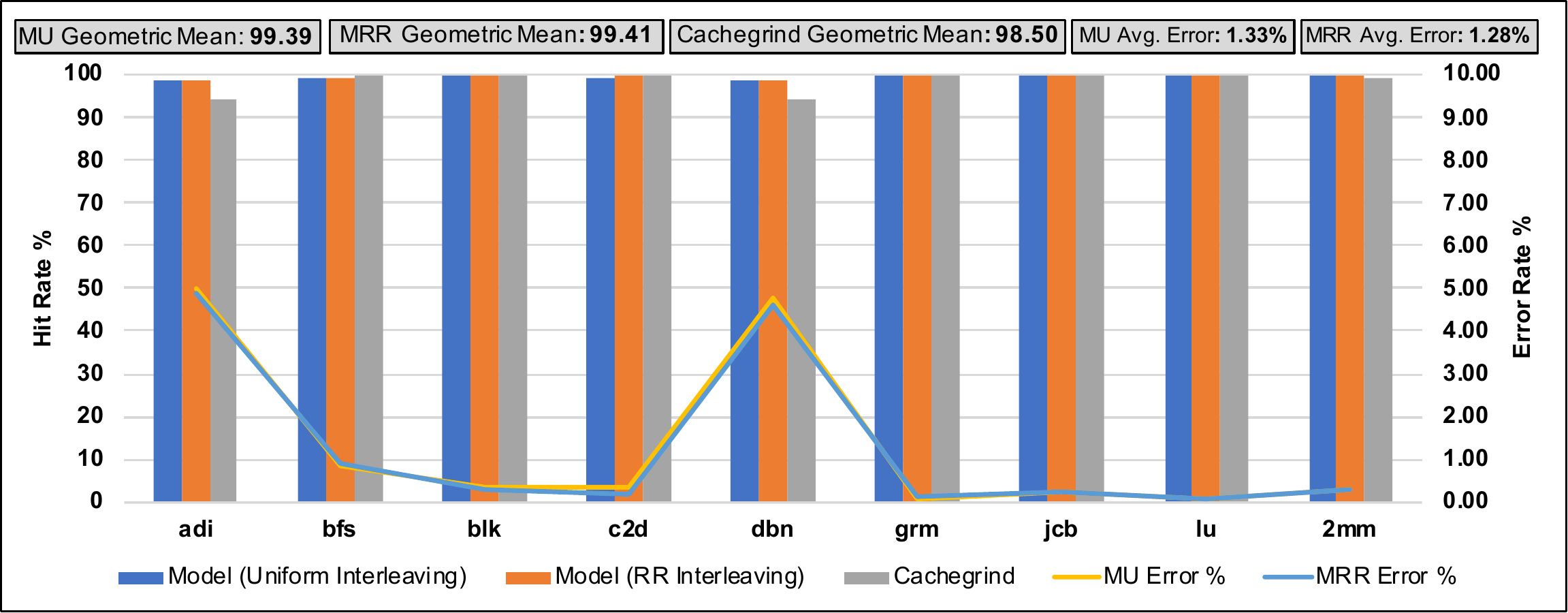}}%
%\vspace{3pt}
\qquad
\subfigure[\textbf{L2 Hit Rates of applications running on 4 core CPU}]{
\label{fig:L2-4Core}%
\includegraphics[width=0.98\linewidth]{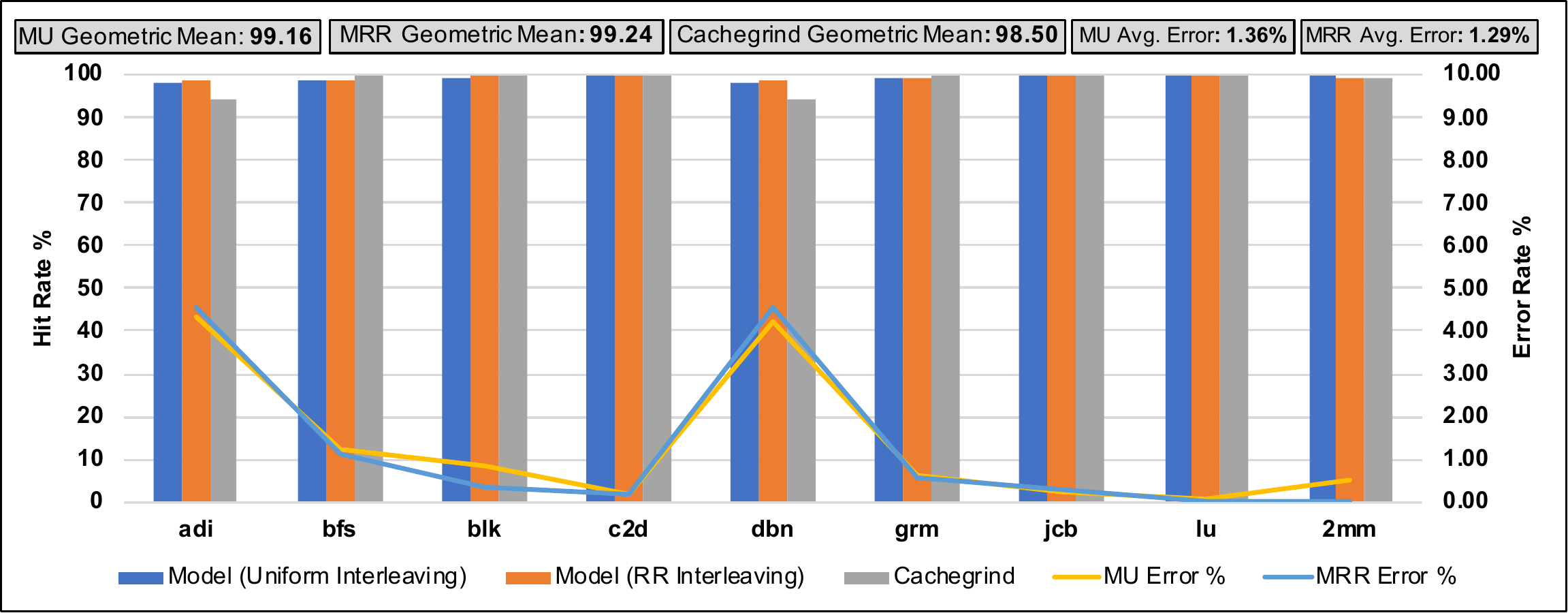}}%
\vspace{1pt}
\qquad
\subfigure[\textbf{L2 Hit Rates of applications running on 8 core CPU}]{
\label{fig:L2-8Core}%
\includegraphics[width=0.98\linewidth]{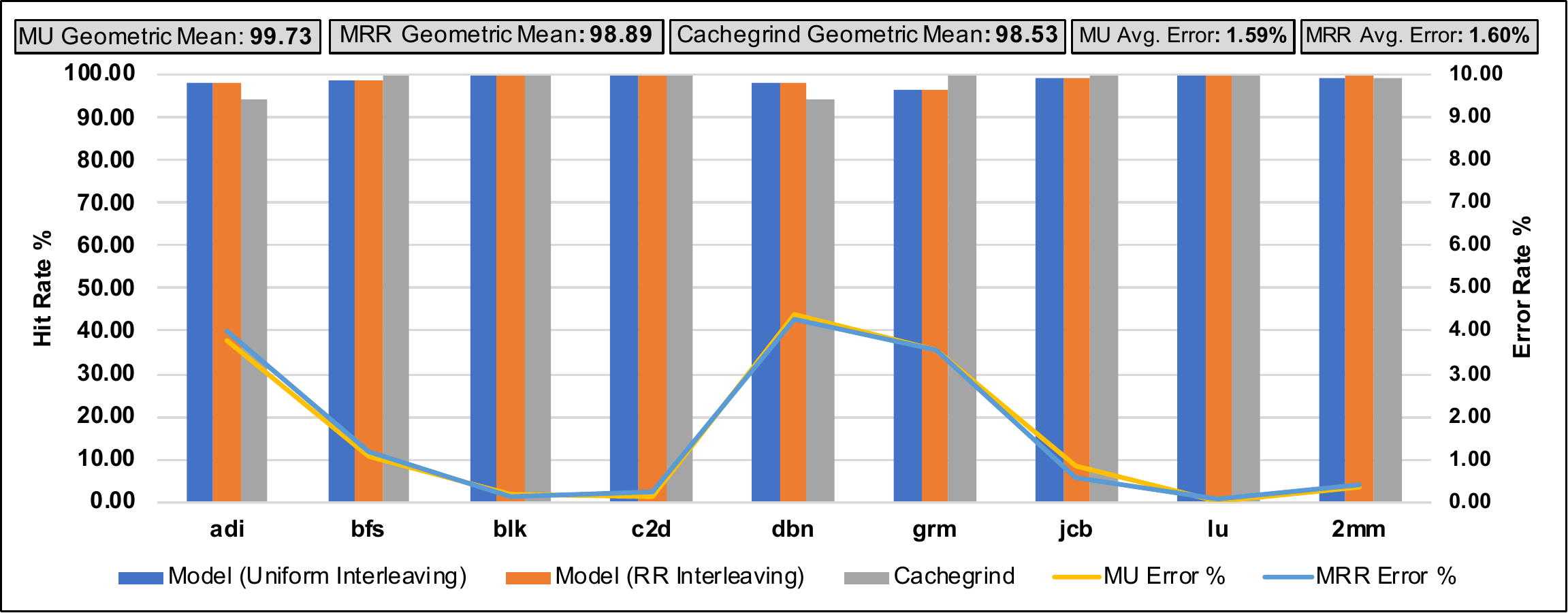}}%
\vspace{3pt}
\qquad
\subfigure[\textbf{L2 Hit Rates of applications running on 16 core CPU}]{
\label{fig:L2-16Core}%
\includegraphics[width=0.98\linewidth]{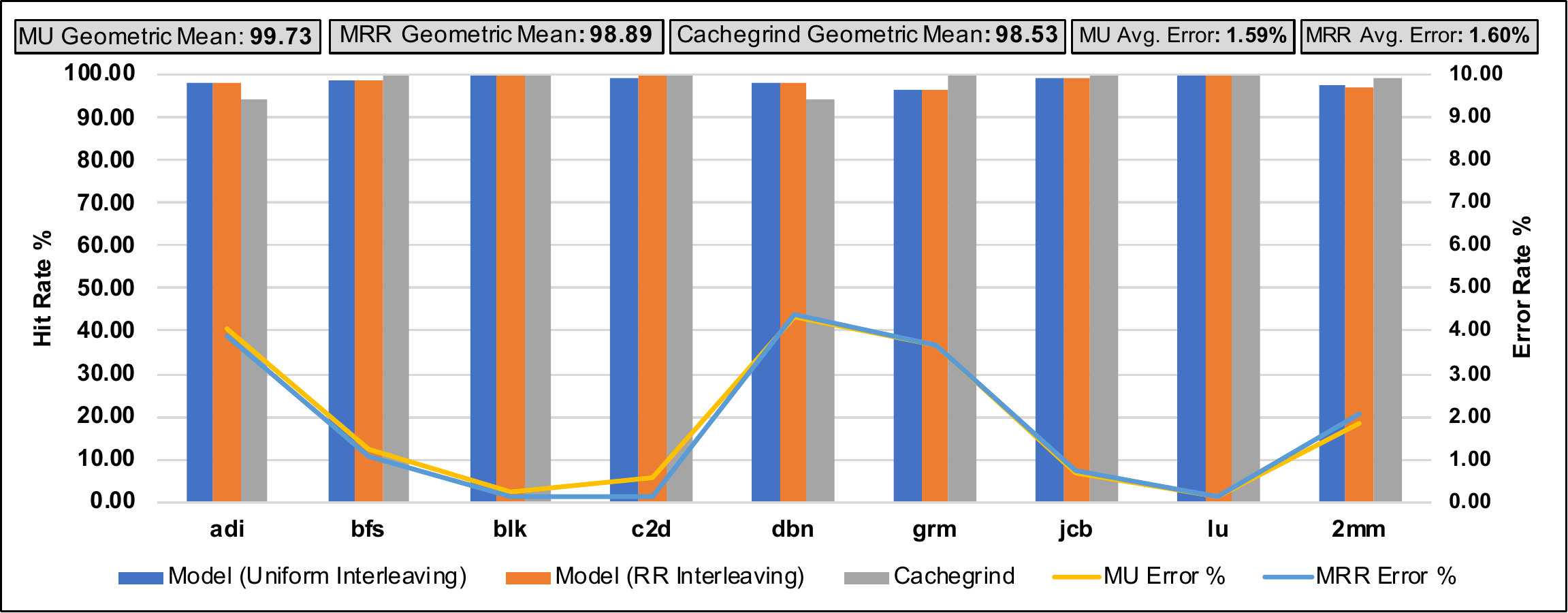}}%
\qquad
\caption{Hit Rate Comparison of Applications Running on CPUs with 128KB Shared L2 Cache}
\label{fig:results-L2}%
\end{figure}

\section{Experimental Results}
In this section, we validate our model and present the results. Table~\ref{table:benchmarks} shows a list of the applications used in the validation. We use nine different applications representing different domains from PolyBench~\cite{polybench}, Rodinia~\cite{rodinia} and PARSEC~\cite{parsec} benchmark suites. For PolyBench, we use the OpenMP implementation by~\cite{polybench-acc}. We choose these benchmark suites as they are widely used for validating performance models. The generated memory trace sizes are also shown for the input used for each application.

We use the Cachegrind tool within the widely used Valgrind~\cite{valgrind} to collect the cache hit rates for different cache configurations. Cachegrind is a dynamic binary analysis tool that performs a trace-driven simulation of a machine's cache as a program executes. The simulated has a split L1 and a unified L2 cache with a write-allocate policy.  The L2 cache is inclusive. Cachegrind does not account for interference from the kernel or other processes when it simulates the caches. It is suitable for verifying our model as we try to evaluate the cache performance of the benchmark applications' standalone execution. It also does not account for virtual to physical address mapping. These properties make it an excellent choice to evaluate our method. We consider two cache levels where the L1 cache is private to each core, and L2 is shared among all the cores. The cache configurations used are as follows.

\begin{itemize}
    \item \textbf{L1 D-Cache} \textit{Size:} 8 KB, \textit{Associativity:} 8, \textit{Line Size:} 64 B
    \item \textbf{L2 Cache} \textit{Size:} 128 KB, \textit{Associativity:} 16, \textit{Line Size:} 64 B
\end{itemize}

% \textcolor{red}{Although modern microprocessors have much larger shared caches, we chose this size so that our benchmark applications does not entirely fit into the cache.}

% \newcolumntype{P}[1]{>{\centering\arraybackslash}p{#1}} %Line 250 defines this

Figure~\ref{fig:results-L1} shows the comparison of hit rates of private L1 cache for each core configuration. We show the hit rates of the applications running on 1, 2, 4, 8, and 16 cores in Figures~\ref{fig:L1-1Core}, \ref{fig:L1-2Core}, \ref{fig:L1-4Core},~\ref{fig:L1-8Core}, and \ref{fig:L1-16Core} respectively. We show the geometric mean of hit rates from our model and Cachegrind in the figures. Our model’s average error rates are 2.18\%, 2.16\%, 2.16\%, 2.13\%, and 1.99\% for the core configurations. We compute the average hit rate of private L1 caches obtained using our model. We change the number of threads/cores using \textit{$OMP\_NUM\_THREADS$} environment variable when we collect the hit rates using Cachegrind. The results show that our model predicts the hit rates of L1 cache accurately with an overall average error rate of \textbf{2.12\%}. As we mimic the memory trace of multi-threaded execution from single-threaded execution, we do not consider the effect of cache coherence in our model. Still, the experiments show promising results.

Figure~\ref{fig:results-L2} compares the hit rates of shared L2 cache for each core configuration. We show the results for both uniform random and round-robin interleaving. We denote them as \textbf{MU} and \textbf{MRR} in the figures. Note that, in the results of a single-core configuration shown in Figure~\ref{fig:L2-1Core}, there is no interleaving. We also show the geometric mean of the hit rates on the L2 cache obtained using our model and Cachegrind. For the single-core configuration, the average error rate is 1.41\% for all the applications. Figures~\ref{fig:L2-2Core}, \ref{fig:L2-4Core}, \ref{fig:L2-8Core} and \ref{fig:L2-16Core} show hit rates for core configurations of 2, 4, 8, and 16 respectively. For uniform random interleaving, average error rates are 1.33\%, 1.36\%, 1.59\%, and 1.85\%, respectively. These make the overall error rate for uniform interleaving \textbf{1.53\%}. For round-robin interleaving of memory traces, average error rates are 1.28\%, 1.29\%, 1.60\%, and 1.81\% respectively for 2, 4, 8, and 16 core configurations. These make the overall average error rate for round-robin interleaving \textbf{1.50\%}.

Overall, our error rates appear tolerable. However, we note quite a bit of difference between the different applications that we tested. In particular the \emph{adi}, \emph{grm}, and \emph{dbn} applications give us trouble when predicting L1 cache hit rates almost independent of core count, whereas only \emph{adi}, and \emph{dbn} show higher fault rates on the shared L2 cache at lower core count. Once we move to a larger core count at L2, the model again starts to over-predict hit rates for \emph{grm}.

The over-predictions in the three applications (adi, grm, and dbn) have two reasons depending on the cache (private or shared) model. For private caches, the discrepancies are because our model works with more instances of thread-specific basic block trace instances than the required. Similarly, the shared cache over-predictions are due to the creation of extra basic block traces during interleaving strategies' mimicking behavior. Overall, although we over-predict some of these applications, we observe low error rates, which can be tolerated concerning the ground truth from Cachegrind.

\section{Related works}
Reuse distance~\cite{Mattson:RD:IBM} analysis has been widely used to predict cache performance~\cite{performance:Beyls:RD:2001, performance:CaBetacaval:2003:ECM,performance:Sen:2013:ROM,multicore:Fast_and_Accurate_Exploration:Maeda}, make policies for cache management~\cite{C.Management:Duong:2012:ICM,C.Management:Keramidas:2007,Das:CacheReplacement} and to predict program locality~\cite{Berg:SS,locality:Ding:2003:PWL,locality:Zhong:2009:PLA,Jiang:RD-Applicable-on-chip}. Researchers also tried to speed up reuse distance calculation by parallelizing the algorithm~\cite{PARDA:Niu} and proposing analytical model and sampling techniques~\cite{Shen:2007:LAU,ppt-amm,chennupati:pads,chennupati:pmbs}. Recently, several research works have been done on reuse distance analysis on multicore processors~\cite{Jiang:RD-Applicable-on-chip,Multicore_Reuse_Analytical:Jasmine,Schuff:2010:AMR:1854273.1854286,multicore:stat_multiprocessor_cache:Berg,Multicore-Aware-Derek,Wu-multicore-journal} and GPUS~\cite{arafa_ics,arafa_ipccc}.

Jiang et al.~\cite{Jiang:RD-Applicable-on-chip} introduced CRD profiles for multicores and provided a probabilistic model to estimate CRDs from the data locality of each thread. They do not consider invalidation for data locality analysis of private caches.

Wu et al.~\cite{Wu-multicore-journal} explored PRD and CRD profiles for performance prediction of loop-based parallel programs. They provided a detailed analysis of the effect of core count on PDR and CRD profiles. They also developed a model for predicting PRD and CRD profiles with core count scaling. The predict the CRD profile with about 90\% accuracy.

Jasmine et al.~\cite{Multicore_Reuse_Analytical:Jasmine} proposed a probabilistic method to calculate the CRD profile of threads sharing a cache and derived coherent reuse profile of each thread considering the effect of cache coherence. They derived the concurrent reuse distance (CRD) profile of each thread, sharing the cache with other threads from the thread's private reuse profile.

Schuff et al.~\cite{Multicore-Aware-Derek} explored reuse distance analysis for shared cache accounting inter-core cache sharing. They also studied PRD profiles considering invalidation-based cache-coherence. They further extended their work to accelerate CRD profile measurement by introducing sampling and parallelization~\cite{Schuff:2010:AMR:1854273.1854286}.

Ding et al.~\cite{Multicore:ding2009a} explored theories and techniques to measure program interaction on multicore processors and introduced a new footprint theory. They proposed a trace-based model that computes a set of per-thread metrics. They compute these metrics by single pass over a concurrent execution of a parallel program. Using these metrics, they propose a scalable per-thread data-sharing model. They also propose an irregular thread interleaving model integrated with the data-sharing model.

Kaxiras et al.~\cite{multicore:Kaxiras} proposed statistical techniques from epidemiological screening and polygraph testing for coherence communication prediction in shared-Memory multiprocessors.

Almost all of these approaches collect traces at different cache levels from parallel execution of the application. Our approach is different since we collect a trace only once from a sequential execution of the application. This makes our approach very scalable with core count.

\section{Conclusion}
Reuse distance analysis has been a valuable tool for application performance prediction. This paper extends reuse distance analysis to the parallel application domain by accounting for inter-thread interactions for shared caches in a static way. It statically predicts the hit rates of a parallel application on private and shared caches from memory traces of the sequential execution of a single-threaded version of the application. This makes the methodology scalable with core counts and cache sizes.
The results show that our model is very accurate for a parallel application's cache hit rate prediction with accuracy ranging from 97.82\% to 98.72\%. We explore various scheduling strategies of OpenMP with different interleaving strategies using our model. Furthermore, the model takes the cache configuration parameters as input, making it suitable for design space exploration and cache sensitivity analysis.

\begin{acks}
The authors would like to thank the reviewers for their feedback. We would also like to thank New Mexico Consortium (NMC) for their continued support and for giving us access to their machines. The authors would also like to thank Dr. David Newsom for donating several machines to the PEARL laboratory at NMSU. Some of the experiments in this paper were run on the donated machines. This work is partially supported by Triad National Security, LLC subcontract \#581326. Parts of this research used resources provided at the Los Alamos National Laboratory Institutional Computing Program. Computations were run on Darwin, a research computing heterogeneous cluster. Any opinions, findings, and/or conclusions expressed in this paper do not necessarily represent the DOE or the U.S. Government's views.
\end{acks}

%\bibliographystyle{ACM-Reference-Format}
%\bibliography{references}

% \appendix

\end{document}